\documentclass[aps,pra,twocolumn,showpacs]{revtex4-1}
\usepackage{graphicx}
\usepackage{amssymb}
\usepackage{amsmath}
\usepackage{bm}
\usepackage{multirow}

\begin{document}

\preprint{}


\title{Observation of a strongly ferromagnetic spinor Bose-Einstein condensate}
\author{SeungJung Huh, Kyungtae Kim, Kiryang Kwon, and Jae-yoon Choi}

\email[]{jaeyoon.choi@kaist.ac.kr}
\affiliation{Department of Physics, Korea Advanced Institute of Science and Technology, Daejeon 34141, Korea }

\date{\today}

\begin{abstract}
We report the observation of strongly ferromagnetic $F=1$ spinor Bose-Einstein condensates of $^7$Li atoms. The condensates are generated in an optical dipole trap without using magnetic Feshbach resonances, so that the condensates have internal spin degrees of freedom.
Studying the non-equilibrium spin dynamics, we have measured the ferromagnetic spin interaction energy and determined the $s$-wave scattering length difference among total spin $f$ channels to be $a_{f=2}-a_{f=0} =-18(3)$ Bohr radius. This strong collision-channel dependence leads to a large variation in the condensate size with different spin composition. We were able to excite a radial monopole mode after a spin-flip transition between the $|m_F=0\rangle$ and $|m_F=1\rangle$ spin states. From the experiments, we estimated the scattering length ratio $a_{f=2}/a_{f=0}=0.27(6)$, and determined $a_{f=2}$ = 7(2) and $a_{f=0}$ = 25(5) Bohr radii, respectively. The results indicate the spin-dependent interaction energy of our system is as large as 46$\%$ of the condensate chemical potential.
\end{abstract}
\maketitle

\section{Introduction}
%
The spinor Bose gas of ultracold atoms has been a pristine platform for studying multi-component superfluid systems like He-3~\cite{Vollhardt1990} and exotic superconductors~\cite{Kruchinin2010}. In such systems the condensate wavefunction has additional spin degrees of freedom and is described by a vector or tensor order parameter~\cite{Ho1998,Ohmi1998}. The system hosts intriguing many-body phases with various topological excitations~\cite{Kawaguchi2012,Stamper2013}, and has been a testbed for quantum information science using the spin squeezed state~\cite{Stamper2013,Pezze2018}. To date, most of the experiments for spin-1 atoms have been carried out using $^{87}$Rb and $^{23}$Na atoms, which exhibit very weak spin-dependent interactions compared to the spin-independent one (e.g., $\sim0.48\%$ in $^{87}$Rb~\cite{Chang2005} and $\sim1.5\%$ in $^{23}$Na atoms~\cite{Black2007}). As a result, even though numerous efforts have been made to tune the scattering length and the spin interaction energy by either using optical~\cite{Hamley2009} or microwave-induced Feshbach resonance~\cite{Zhang2009,Papoular2010}, spinor condensates with strong interactions have been largely unexplored. 

Spinor Bose-Einstein condensates of $^{7}$Li are predicted to display strong spin-dependent interaction ($\sim45\%$ of the density-density interaction~\cite{Stamper2013}) and they have recently attracted increasing attention. The strong ferromagnetic spin interactions provide new opportunities to investigate the complex interplays between magnetic order and superfluidity~\cite{Sonin2010,Fang2016,Armaitis2017,Sonin2018,Kim2017,Fava2018}, universal coarsening dynamics after a quantum phase transition~\cite{Barnett2011,Williamson2016,Williamson2017,Fujimoto2018}, and to explore rich phases in optical lattices~\cite{Rodriguez2011,So2017}. In addition, the strong interactions can speed up the one-axis squeezing dynamics~\cite{Kitagawa1993,Sorenson2001,Riedel2020,Gross2011,Hamley2012}, and macroscopic superposition states might be generated within feasible time scales~\cite{Micheli2003}. However, experimental studies of spinor condensates of $^7$Li atoms have been limited by the incompatible experimental conditions needed to produce condensates with spin degrees of freedom. Since the scattering length of the $^7$Li is very small under moderate magnetic fields, the condensates have been produced under a strong magnetic field ($\sim$700~G)~\cite{pollack2009,Gross2008,Dimitrova2017,Kim2019}, where the scattering length is increased by using  Feshbach resonances~\cite{Chin2010}. The strong Feshbach field, however, induces a large quadratic Zeeman shift between magnetic sub-levels, freezing out the spin-mixing collision process. This is in stark contrast to $^{87}$Rb and $^{23}$Na atoms, where the scattering lengths are large enough to generate Bose-Einstein condensates in an optical dipole trap under a residual magnetic field~\cite{Chang2005,Stamper1999}. 

In this work, we have overcome the technical difficulty and report the creation of strongly ferromagnetic $F=1$ spinor Bose-Einstein condensates. By preparing a thermal gas near the quantum degeneracy in a large volume optical trap, pure condensates containing up to $N=7\times10^5$ atoms have been produced after evaporation cooling without the aid of Feshbach resonance. To measure the spin-dependent interaction energy, we performed two independent experiments. We investigated magnetization dynamics across the quantum phase transition point (between a polar (P) phase and an easy-plane ferromagnetic (EPF) phase)~\cite{Sadler2006}, and examined the coherent spin-mixing dynamics under various magnetic fields~\cite{Chang2005}. In both experiments, we obtained similar results for the scattering length differences, $a_{f=2}-a_{f=0}=-18(3)~a_{\text{B}}$, where $a_{\text{B}}=52.9$~pm is the Bohr radius. Such a large difference in scattering length caused noticeable changes in the condensate size with different spin state. A two-dimensional breathing mode was excited after the spin-flip transition from the $|m_F=0\rangle$ to the $|m_F=1\rangle$ state. From this experiment, we were able to obtain a scattering length ratio, $a_{f=2}/a_{f=0}=0.27(6)$, and determine $a_{f=2}$ = 7(2) and $a_{f=0}$ = 25(5)~$a_{\text{B}}$, respectively. Our results are close to the theoretical estimation obtained from molecular energy level calculations ($a_{f=2}$ = 6.8~$a_{\text{B}}$ and $a_{f=0}$ = 23.9~$a_{\text{B}}$)~\cite{Stamper2013,Julienne2014}, and imply strongly ferromagnetic interactions among the $^7$Li atoms.

 This paper is structured as follows. In Sec.~II, the theoretical backgrounds for measuring scattering length difference are introduced. In Sec.~III, we present the experimental setup and cooling process used to realize the Bose-Einstein condensates under a few Gauss of magnetic field. In Sec.~IV, the two different experiments are introduced, which measure the scattering length difference and the ratio among the allowed spin channels. We provide a summary and outlook in Sec.~V.

\section{Measuring spin-dependent interaction energy}

We consider a Hamiltonian for a spin-1 Bose-Einstein condensate in a homogenous magnetic field $B$, 
\begin{eqnarray}
H&=&\int d\mathbf{r}~\hat{\Psi}^{\dagger}(\mathbf{r}) \left( -\frac{\hbar^2}{2m}\nabla^2+V(\mathbf{r}) + q\hat{F}_z^2   \right)\hat{\Psi}(\mathbf{r})\nonumber \\ 
&&+\int d\mathbf{r}\left(\frac{1}{2}c_0n^2 +\frac{1}{2}c_2 |\langle \mathbf{F} \rangle|^2 \right),
\end{eqnarray} where $\hat{\Psi}(\mathbf{r})=(\hat{\psi}_1,\hat{\psi}_0,\hat{\psi}_{-1})^{T}$ is a three component bosonic field operator, $m$ is the atomic mass, $\hbar$ is the Planck constant divided by $2\pi$, $V(\mathbf{r})$ is a trapping potential, $\mathbf{F}=(\hat{F}_x,\hat{F}_y,\hat{F}_z)$ is a spin-one matrix operator, and $n=\hat{\Psi}^{\dagger}\hat{\Psi}$ is the condensate density. The linear Zeeman shift is removed because of magnetization conservation, and the quadratic Zeeman shift for $^7$Li is $q=(h\times610$~Hz/G$^2){~B^2}$. The $c_0=4\pi\hbar^2(2a_{f=2}+a_{f=0})/3m$ and  $c_2=4\pi\hbar^2(a_{f=2}-a_{f=0})/3m$ are the spin-independent and spin-dependent interaction coefficients, respectively, where $a_f$ is the $s$-wave scattering length in the total spin~$f$ channel.

The ground state spin structures of the ferromagnetic spinor condensates are determined by the competition between the spin-dependent interaction energy ($c=c_2n$) and the quadratic Zeeman shift ($q$), where the polar phase and the easy-plane ferromagnetic phase are separated by a quantum critical point, $q_c=2|c|$ [Fig.~1(a)]. Below the critical point, the P phase is dynamically unstable, and the quantum fluctuations of the $m_F=\pm1$ spin pairs are amplified to form ferromagnetic spin domains~\cite{Zhang2005b,Saito2005,Saito2007}. The magnetization of the P phase across the critical point has been previously observed for $^{87}$Rb atoms~\cite{Sadler2006,Anquez2016}, and in this work, we will associate the spin interaction coefficient with the quadratic Zeeman energy by locating the quantum critical point ($|c_2|=q_c/2n$) and determine the scattering length difference.

\begin{figure}
\includegraphics[width=0.8\columnwidth]{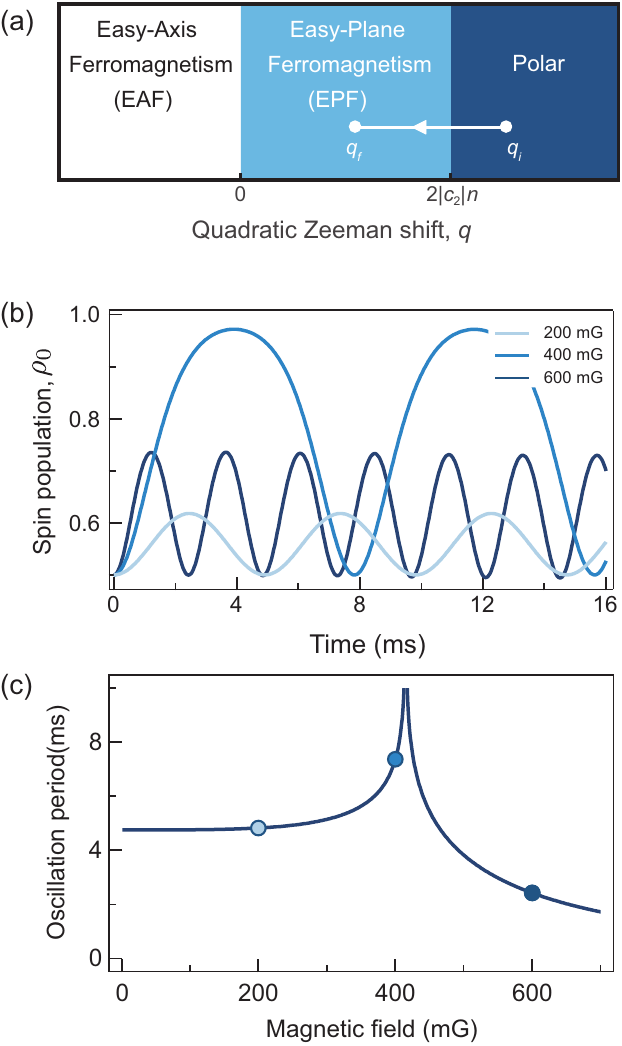}
\caption{(a) Ground state phase diagram of a ferromagnetic spinor condensate. As the quadratic Zeeman energy is varied, the polar phase undergoes a phase transition to the easy-plan ferromagnetic (EPF) phase ($0<q<q_c=2|c_2|n$) or to the easy-axis ferromagnetic (EAF) phase ($q<0$). (b), (c)  Analytic calculation of spinor dynamics under various magnetic fields with single mode approximation. The initial spin state is $\boldsymbol{\zeta}=(1/2,1/\sqrt{2},1/2)$, and the spin interaction energy is set to $c=-h\times 105$~Hz. The oscillation periods shows a non-linear dependence on magnetic field with a peak at $q(B_d)=|c|$.
\label{Background}}
\end{figure}

The spin-dependent interactions in the Hamiltonian involve a spin-mixing collision process between the magnetic sub-levels, $2|m_F=0\rangle\leftrightarrow|m_F=1\rangle+|m_F=-1\rangle$, leading to coherent oscillation dynamics in the spin population. The spinor dynamics can be simplified by using a single mode approximation (SMA)~\cite{Zhang2005a}, which assumes the condensates with different spin states share the same spatial wavefunction $\phi(\bf{r})$. The condensate order parameter is written as $\phi(\bf{r})\boldsymbol{\zeta}$, a product of $\phi(\bf{r})$ and spinor $\boldsymbol{\zeta}=(\sqrt{\rho_{1}}e^{i\theta_{1}},\sqrt{\rho_{0}}e^{i\theta_{0}},\sqrt{\rho_{-1}}e^{i\theta_{-1}})$, where the $\rho_{m_F}$ and $\theta_{m_F}$ denote the fractional population and phases of the Zeeman sublevels $|m_F\rangle$, respectively. Within the SMA the dynamics are described by two canonical variables, $\rho_0$ and $\theta=\theta_{1}+\theta_{-1}-2\theta_{0}$, at a given magnetization, $M=\rho_{1}-\rho_{-1}$~\cite{Zhang2005a}. For the ferromagnetic spin interaction ($c_2<0$), the dynamics are divided into interaction ($|c|\gg q$) and the Zeeman energy dominant ($|c|\ll q$) regimes, and a singular behavior is expected when these two energy scales are comparable. For example, with an initial state $\rho_0(0)=\alpha$, $\theta(0)=\beta$, and $M=0$, the oscillation period can be computed by elliptic integration of the first kind, and diverges at the magnetic field $B_d$, which satisfies $q(B_d)=|c|\alpha(1+\cos\beta)$  [Fig.~1(c)]. Thus, by investigating the coherent oscillation dynamics under various magnetic fields, the spin-dependent interaction energy and the scattering length difference can be measured~\cite{Chang2005,Black2007}.

\section{Experimental setup}

\begin{table}[]
\begin{tabular}{l|c|c|c}
   & $m_F=1$     & $m_F=0$   & $m_F=-1$\\ \hline
$m_F=1$  & \multicolumn{1}{c|}{$a_2=6.8~a_{\text{B}}$} & \multicolumn{1}{c|}{$a_2$} &  $18.2~a_{\text{B}}$  \\
$m_F=0$  & \multicolumn{1}{c|}{$a_2$}         & \multicolumn{1}{c|}{$\frac{(2a_2+a_0)}{3}=12.5~a_{\text{B}}$} &  $a_2$  \\
$m_F=-1$ & \multicolumn{1}{c|}{$\frac{(a_2+2a_0)}{3}=18.2~a_{\text{B}}$}         & \multicolumn{1}{c|}{$a_2$} &  $a_2$ 
\end{tabular}
\caption{Scattering length for binary $s$-wave collisions among atoms in the $F = 1$ manifold. The estimation refers to the theoretical calculation with $a_{f=2}=6.8~a_{\text{B}}$ and $a_{f=0}=23.9~a_{\text{B}}$~\cite{Stamper2013,Julienne2014}.}
\end{table}

\begin{figure*}
\includegraphics[width=1.9\columnwidth]{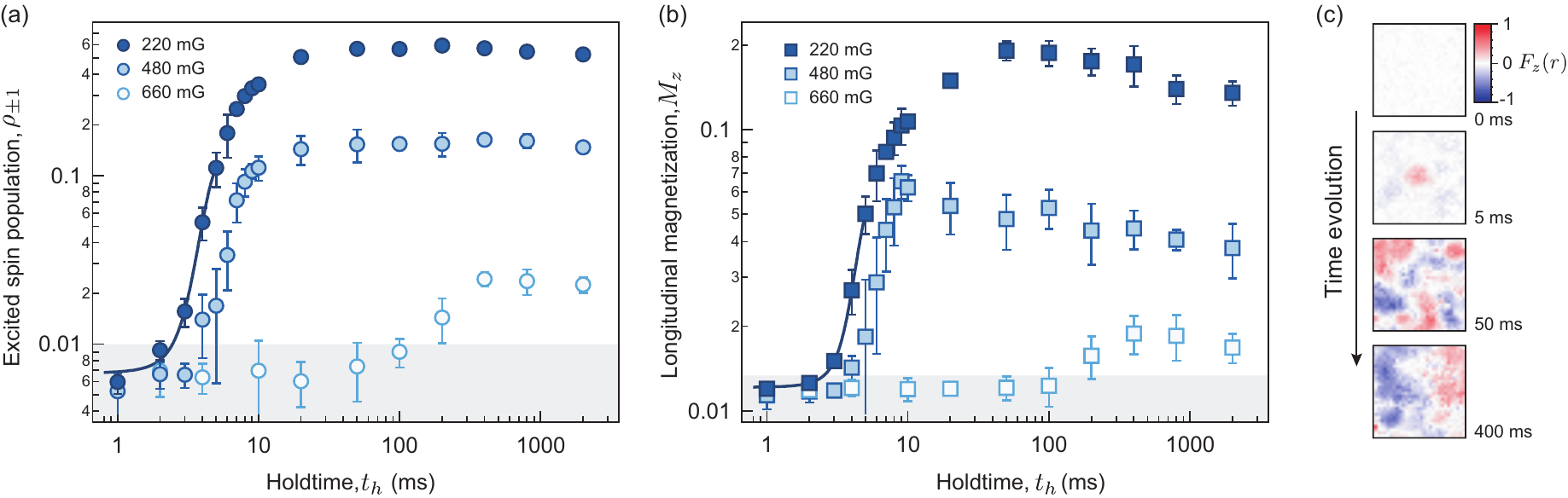}
\caption{Magnetization dynamics of the $^7$Li spinor BEC after the quench. (a) The excitation population $\rho_{\pm1}$ and (b) the mean magnetization $M_z$ as a function of hold time for three different magnetic field ($B_f=220$~mG in dark blue, $B_f=480$~mG in sky blue, and $B_f=660$~mG in light blue). The dynamical instability of the polar phase is represented as exponential increases of the magnetization. Solid line at $B_f=220$~mG is the sigmoid fit with $t_h\leq5$~ms. The data points represent averages of 5 independent experiments, and the error bars denote the one standard deviation. The gray regions mark statistical bounds, where the observables are compatible with zero. (c) Temporal and spatial evolution of the longitudinal magnetization $F_z(r)$ after the quench ($B_f=220$~mG). The instability starts at the condensate center, and the ferromagnetic domains are rapidly developed in $\sim$10~ms. In the long time scale ($\sim$100~ms), the domains are merged and separated because of the residual field gradient. The central field of view used in the analysis is $120~\mu{}$m$\times120~\mu{}$m (appendix B).
\label{Bquench}}
\end{figure*}

 The experiment began by making Bose-Einstein condensates of $^7$Li atoms in an optical dipole trap under a few Gauss of magnetic field. To obtain the condensates, we first performed microwave evaporative cooling of the atoms in the upper hyperfine state $|F=2,m_F=2\rangle$ in an optically plugged quadrupole magnetic trap~\cite{Kim2019}, and then transferred the cold atoms to a quasi-two-dimensional (quasi-2D) optical trap~\cite{Clade2009}. Since the quasi-2D trap has a larger trap volume than the other trap geometries used in previous experiments~\cite{pollack2009,Gross2008,Dimitrova2017,Kim2019}, we are able to trap more atoms with lower entropy per particle at a given trap depth~\cite{Lin2009}. 
 
After loading the cold atoms in the quasi-2D trap, we applied double Landau-Zener sweeps at 20~G to prepare the atoms in the $|F=1,m_F=0\rangle$ state, which has the largest intraspecies scattering lengths among the $F=1$ hyperfine state [Table~1]. The collision rate was high enough ($\sim 150$~Hz with harmonic approximation) for conventional evaporation cooling, and we cooled the atoms by lowering the trap depth. The bimodal distribution of a BEC was observed 0.5~s after the cooling, and the pure condensates containing $N_c=7\times10^5$ atoms were generated after 5~s of full evaporation. During the evaporation process, we kept the magnetic field along the $z$-axis at 20~G to prevent the thermal populations in the other spin states, and lowered the field to a few Gauss after generating the BECs. The final trap frequencies were $(\omega_x, \omega_y, \omega_z) = 2\pi\times (8,10,680)$~Hz. The chemical potential ($\mu=h\times{}320$~Hz) was less than half of the axial trap frequency so that our system satisfied the 2D criterion. The lifetime of the condensates was over 100~s. Details on the experimental parameters and evaporative cooling process are provided in the appendix A.
  
 The atomic density for each spin component ($m_z=\pm1,0$) was detected by an absorption image after Stern-Gerlach spin separation. After switching off the optical trap, the magnetic field was rotated along the $xy$-plane in 3~ms, and a field gradient of 13~G/cm was applied for 5~ms in the $x$-direction. After 8~ms of free expansion, all the spin components were spatially separated. Then, the atoms were optically pumped into the $|F=2\rangle$ state, and we took an absorption imaging using the $|F=2\rangle\rightarrow |F'=3\rangle$ resonant light.

\section{Results}

\subsection{Quenched ferromagnetic Bose gas}

We employed a rapid quench experiment to identify the critical point, and to determine spin-dependent interaction energy and scattering length difference. The BECs of $^7$Li were prepared in the $|m_z=0\rangle~(|0\rangle)$ state under 1~G of magnetic field along the $z$-direction. At this field strength, the quadratic Zeeman shift ($q=h\times610$~Hz) was larger than the critical point $q_{c}$, and the initial state was stable over several seconds. Then, we reduced the magnetic field to $B_f$ in 1.4~ms, and a density profile of each spin state $(n_{1},n_{0},n_{-1})$ was recorded using spin-separated absorption imaging after a variable hold time $t_h$  (absorption images are shown in appendix B). With this imaging technique, we were able to study the stability of the polar phase from the appearance of the $|m_z=\pm1\rangle~(|\pm1\rangle)$ spin components and observe the ferromagnetic spin domains by calculating the average longitudinal magnetization, $M_z(t)=\int |\langle F_z (r,t)\rangle| d^2r/A_R$, where $A_R$ is the area of the central region.

 \begin{figure}
\includegraphics[width=0.85\columnwidth]{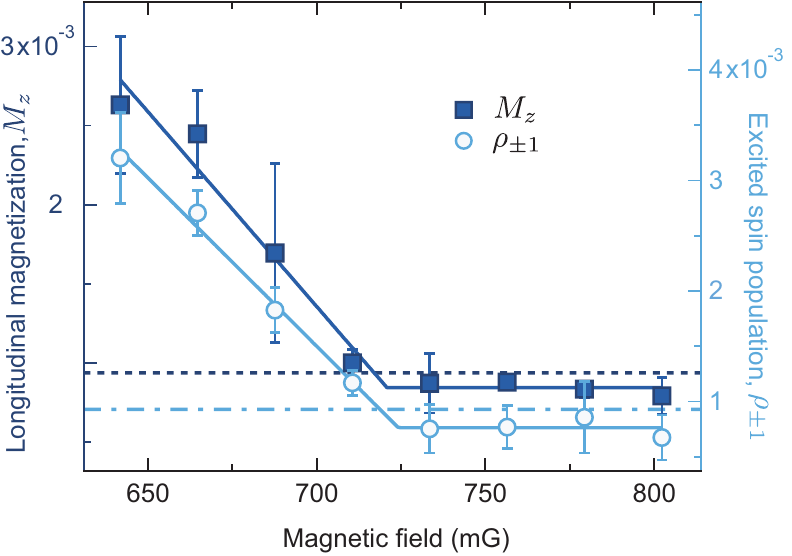}
\caption{Identifying the quantum critical point. The steady state mean magnetization $M_z$ (dark blue) and excited spin population $\rho_{\pm1}$ (light blue) at a constant hold time $t_h^{ss}=500$~ms as a function of magnetic field. The dashed line and dashed-dotted line represent statistical zero for each observable. The plot shows sharp rises of $M_z$ and $\rho_{\pm1}$ below the critical magnetic field, $B_c=720(12)$~mG, which is obtained from the bi-linear fit (solid lines, described in the text). The uncertainty of the critical point indicates 95$\%$ confidence interval for the fit. 
 \label{Bquench}}
\end{figure}

In deep quench ($B_f=220$~mG), the domains first appear at the trap center, where density is high, and develop over the entire sample with the hold time [Fig.~2(c)]. The early time dynamics show exponential growth of $M_z(t)$ with a time constant $\tau_{\text{fm}}$=0.52(3)~ms, which is the characteristic feature of the dynamical instability~\cite{Saito2005,Saito2007}. Since the growth rate is determined by the spin interaction energy, $1/\tau_{\text{fm}}=2|c|/\hbar$, we can roughly estimate the spin interaction energy to be $|c|\sim h\times 150$~Hz. After 40~ms, the excited spin state population $\rho_{\pm1}= (n_{1}+n_{-1})/n_{tot}$ reaches a steady state [Fig.~2(a)], and the magnetization gradually decreases [Fig.~2(b)]. In this regime, the polarized spin domains start to merge, and the number of spin domains decreases, revealing the coarsening dynamics of ferromagnetic spin domains~\cite{Guzman2011,De2014}. After 100~ms, because of the residual field gradient of 2~mG/cm, we observe a phase separation between the $|\pm1\rangle$ spin state along the gradient field direction~\cite{Stenger1998}. When we quenched the magnetic field near the critical point ($B_f=660$~mG), the growth dynamics slowed down (a steady state was achieved after a few hundreds of ms) with a few percent of the population in the $|\pm1\rangle$ spin state.

To determine the quantum critical point, we measured the $M_z$ and excitation population $\rho_{\pm1}$ at a fixed hold time $t_h^{ss}=$500~ms, and scanned the magnetic field. As shown in Fig.~3, a clear onset of the $M_z$ ($\rho_{\pm1}$) is observed below the critical magnetic field $B_c$, which can be extracted from a simple double linear fit, $M_z(B)=M_0+M_1\cdot\max[(B_c-B),0]$, where $B_c=720(12)$~mG and $q_c=320(12)$~Hz. In the vicinity of the critical point, the spin domains appear only at the trap center, such that the spin-dependent interaction energy can be related to the critical point as $|c_2|n_c=q_c/2=160(6)$~Hz, where $n_c$ is the condensate peak density. The peak density was obtained by measuring the mean field energy of the condensate, and the scattering length of the polar condensates was assumed to $12.5~a_{\text{B}}$ [Table~1]. During the measurement, we applied short pulse (10~$\mu$s) laser light with low intensity ($0.1~I_{\text{s}}$) to minimize doppler shift and de-pumping effect. The saturation intensity $I_{\text{s}}$ of the $^7$Li atoms is 2.54~mW/cm$^2$. The peak density was $n_c=2.9(5)\times10^{13}$/cm$^{3}$, and we calculated $a_{f=2}-a_{f=0}=-18(3)~a_{\text{B}}$. This result is close to the theoretical calculation obtained from the molecular levels, $a_{f=2}-a_{f=0}=-17.1~a_{\text{B}}$~\cite{Stamper2013,Julienne2014}.

\subsection{Coherent spin-mixing dynamics}

As a complementary experiment, we studied the coherent spin-mixing dynamics under a various magnetic fields to measure the spin-dependent interaction energy and the $s$-wave scattering length difference. We took the spinor vector $\boldsymbol{\zeta_0}=(1/2,1/\sqrt{2},1/2)$ as an initial state, where  the single mode approximation (SMA) predicts a divergence of the oscillation period and amplitude under an external magnetic field $B_d$ [Fig.~1(c)]. The initial state was prepared by applying an RF-pulse to the BECs in the $|m_F=1\rangle$ spin state. Here, the condensates were produced after the evaporation cooling in the optical trap, but after a longer cooling time (8~s) because of its smaller scattering length compared to the $|m_F=0\rangle$ spin state [Table~1]. The magnetic field was stabilized before applying the rotating pulse, so that the spin dynamics started right after the RF-pulse. Each spin component was resolved by absorption imaging after the gradient pulse during the time-of-flight.

\begin{figure}
\includegraphics[width=0.85\columnwidth]{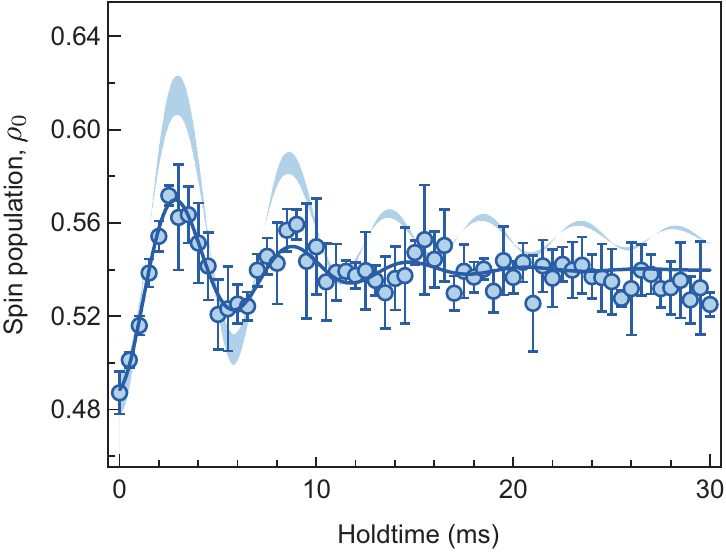}
\caption{Coherent spinor dynamics in $^7$Li condensates. Time evolution of the relative spin population $\rho_0$ at magnetic field 0.17~G with the initial state $\rho_0(0)=0.48$, $\theta(0)=0$, and $M=0$. Solid line is the characteristic damped sinusoidal fit and the shaded envelopes represent the LDA (described in the text) with experimental uncertainties: the initial state spin composition, the condensate density, and the magnetic field. The spin domains are developed after one oscillation ($\sim 6$~ms), reducing the oscillation amplitude. The LDA curve shows the mean spin-dependent interaction $c_2n_{\text{avg}}=-h\times 89(2)$~Hz. The central area in the analysis is $80~\mu{}$m$\times 80~\mu{}$m.
\label{SMD}}
\end{figure}

Fig.~4 displays the time evolution of the relative population in the $|0\rangle$ spin state $\rho_0(t)$ under constant magnetic field. The coherent oscillation represents the spin-mixing dynamics between the magnetic sub-levels, and the total magnetization is preserved during the dynamics. The short oscillation period ($\sim6$~ms) implies a strong spin-dependent interaction energy, and the initial increase in the spin population $\rho_0(t)$ indicates the dynamics are driven by the ferromagnetic spin interaction ($c_2<0$). 

Since the oscillation amplitude rapidly decreases over the hold time, we characterized the spinor dynamics using the damped sinusoidal function, $\rho_0(t)=\rho_{ss}+\rho_{A}e^{-t/\tau}\sin(\omega t+\phi)$. $\rho_{ss}$ is the steady-state value, $\tau$ is a damping constant, $\omega$ is oscillation frequency, $\rho_{A}$ is the oscillation amplitude, and $\phi$ is the phase set by the initial state $\rho_0(0)\simeq0.5$. Such strong damping can be understood in the context of the break down of the single mode approximation. Our condensate size, $(R_{x},R_{y})=(100,80)~\mu$m, is much larger than the spin healing length $\xi_{s}=h/\sqrt{2m|c|}=18~\mu{}$m. In this regime, the coherence of the spinor dynamics can be lost for the following reasons. First, as in the quenched experiment, the dynamical instability amplifies spin fluctuations of the initial state, generating multiple spin domains in a random position. Second, when the condensates have an inhomogeneous density distribution, for example, BECs in a harmonic potential, the spin interaction energy and the oscillation frequency have spatial dependence. In both cases, the coherent dynamics can be dephased after averaging the central area of the condensates, which is necessary to reach a sufficient signal-to-noise ratio. 

\begin{figure}
\includegraphics[width=0.85\columnwidth]{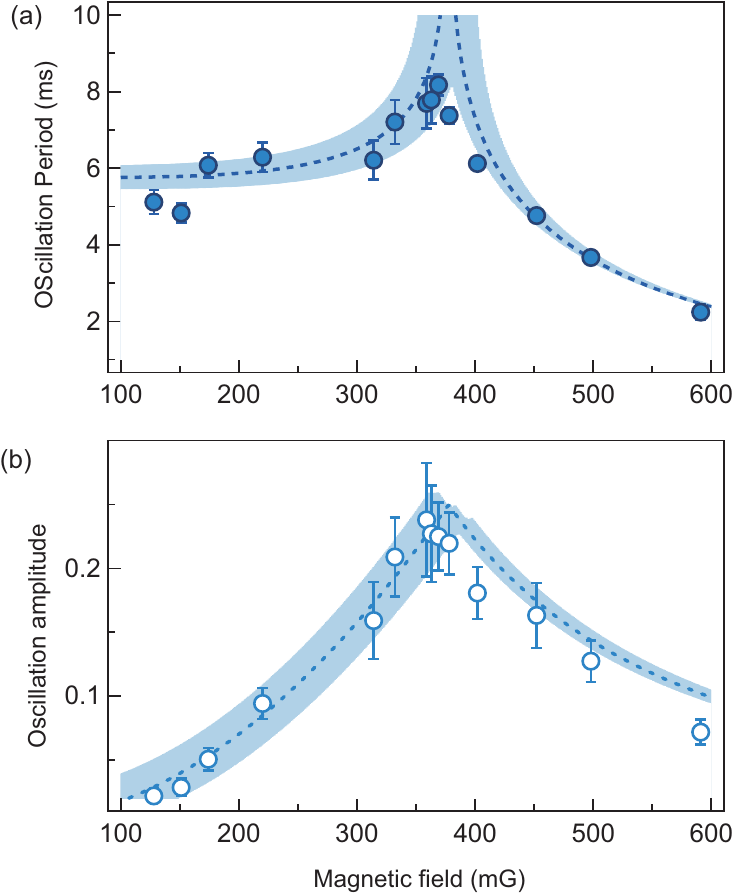}
\caption{The oscillation period (a) and amplitude (b) as a function of magnetic field with the initial spinor vector $\bf{\zeta}_0$. Dashed lines are the single mode theory fittings with the initial spinor vector $\boldsymbol\zeta_0$ ($\rho_0(0)=0.5$, $\theta(0)=0$, and $M=0$). The shaded regions include the statistical uncertainties in the experiments. The peak values are located near the magnetic field $B_d\sim0.4$~G. The data points are the mean value of 5 independent experiments, and the error bars represent the 95$\%$ confidence interval of the damped oscillation fits. 
\label{SMDsummary}}
\end{figure}

Despite the strong damping in our experiment, we note that the characteristic fit function reflects the key features of the single mode spinor dynamics. This has been pointed out in the numerical study~\cite{MurPetit2009}, which shows that the initial suppression of the coherent spin dynamics can be mostly attributed to the inhomogeneous density. In the study, the magnetization from the dynamical instability is negligible in the first few oscillation cycles, and thus the short time dynamics can be approximated to the damped harmonic function, where its oscillation amplitude and the frequency are well matched to those of single mode spinor dynamics with average density ($n_{\text{avg}}$). 
 
In the experiment, the effect of density inhomogeneity is represented by increases in the damping coefficient as the region of interest grows larger, which leads us to take the local density approximation (LDA): the spin population $\rho_0(r,t)$ evolves in a single spatial mode with the local spin interaction energy, $c_2n(r)$. The LDA can well describe the short time dynamics with almost the same oscillation period [Fig.~4]. The smaller amplitude observed in the experiment might be attributed to the spin domains. We further studied the coherent spin dynamics under various magnetic fields and summarize the results in Fig.~5. The oscillation period and its amplitude show a peak near $B_d\sim0.4$~G, which is the characteristic feature of the simple theory with ferromagnetic spin interaction. Treating the spin-dependent interaction energy ($c=c_2n_{\text{avg}}$) as a free parameter in the SMA, we obtained the best fit using $c=-h\times$88(2)~Hz. The average density was calculated from the mean field energy, $n_{\text{avg}}=1.5(3)\times10^{13}$~cm$^{-3}$, and we determined the scattering length difference to be $a_{f=2}-a_{f=0}= -18(3)~a_{\text{B}}$, which is similar to the results of previous section.

\subsection{The scattering length ratio, ${a_{f=2}/a_{f=0}}$}

Lastly, we measured the scattering length ratio between the total spin channel in the binary collision, $\gamma=a_{f=2}/a_{f=0}$, such that the scattering length for each collision channel can be determined together with the previous results. We first estimated the ratio by directly comparing the radial size of the trapped condensate with different spin states ($|m_F=0\rangle$ and $|m_F=1\rangle$). Pure condensates of both spin states were prepared by evaporation cooling. More atoms were in the $|m_F=0\rangle$ state after full evaporation ($N_{0}=6.8\times 10^5$ and $N_{1}=4.4\times 10^5$). Fig.~6(a) and (b) show the condensate in the $xy$-plane and its horizontal cross section for both spin states. From the radial scaling in two dimensions, $R_{m_F}\propto({a^s_{m_F}}N_{m_F})^{1/4}$, where $a^s_{m_F}$ is the scattering length between two spin-$m_F$ atoms, we have $a^s_{m_F=0}/a^s_{m_F=1}=1.8(3)$ and $\gamma=0.29(4)$.

The results signify that a radial breathing mode can be excited after a spin-flip transition because of the sudden change in the scattering length. For example, after the spin-flip transition from $|m_F=0\rangle$ to $|m_F=1\rangle$ state, the condensate will shrink after the pulse and be compressed until the initial potential energy is fully converted into interaction energy. Afterwards, it expands, displaying a radial oscillation. The radial breathing mode is of particular interest in an isotropic two dimensional harmonic potential, where the system has a dynamical symmetry described by the Lorentz group SO(2,1)~\cite{Pitaevskii1997,Saint2019}. Under the dynamical symmetry, the superfluid dynamics are greatly simplified. For example, solving the 2D Gross-Pitaevskii equation, the potential energy per particle $E_{\text{pot}}(t)$ evolves as
\begin{equation}
E_{\text{pot}}(t) =\frac{1}{2}(\Delta{}E\cos(\omega_B{}t)+E_{\text{tot}}),
\end{equation} where $E_{\text{tot}}$ is the total energy of the system including interaction energy ($E_{\text{int}}$), potential energy ($E_{\text{pot}}$), and kinetic energy ($E_{\text{kin}}$). The $\Delta{}E=[E_{\text{pot}}(0)-E_{\text{int}}(0)-E_{\text{kin}}(0)]$ is oscillation amplitude, and $\omega_B$ is the breathing mode frequency. In the Thomas-Fermi approximation, the two-dimensional potential energy is equal to the interaction energy, $E_{\text{pot}}(0)=E_{\text{int}}^{0}$, so that $\Delta{}E=E_{\text{int}}^{0}-E_{\text{int}}^{1}$ and ${E_{\text{tot}}}=E_{\text{int}}^{0}+E_{\text{int}}^{1}$. The $E_{\text{int}}^{m_F}$ is the mean field interaction energy of the condensates in the $|m_F\rangle$ spin state. Since the interaction energy is proportional to the $a^s_{m_F}$, the normalized oscillation amplitude, $\Delta E/2E_{\text{pot}}(0)$, and the offset, $E_{\text{tot}}/2E_{\text{pot}}(0)$, can be expressed as a function of the scattering length ratio $\gamma$,   
\begin{eqnarray}
\frac{\Delta E}{2E_{\text{pot}}(0)} &=& \frac{1-\gamma}{2+4\gamma},\\
\frac{E_{\text{tot}}}{2E_{\text{pot}}(0)} &=& \frac{1+5\gamma}{2+4\gamma}.
\end{eqnarray} Therefore, the microscopic changes in scattering length can be inferred by studying the 2D breathing mode dynamics.

 \begin{figure}
\includegraphics[width=0.9\columnwidth]{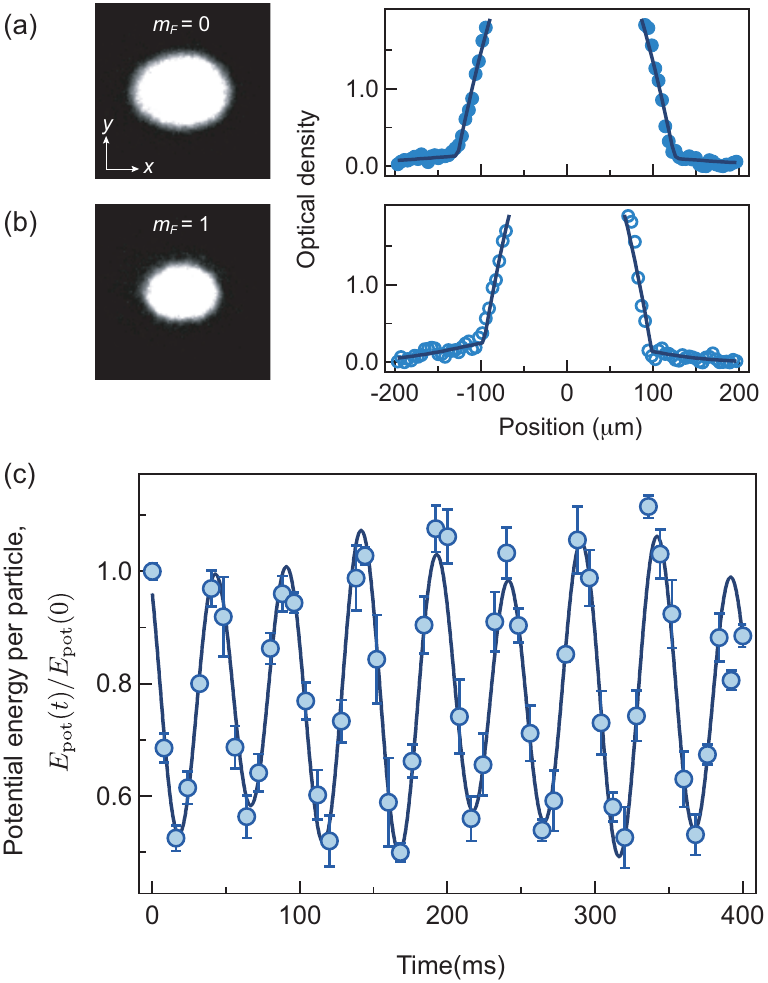}
\caption{In-trap images and its cross section views of the condensates with (a) $|m_F=0\rangle$ and (b) $|m_F=1\rangle$ spin states. Solid lines are the Thomas-Fermi fits with background thermal distribution. (c) Time evolution of the potential energy per particle $E_{\text{pot}}$ in two dimensional harmonic trap. The potential energy is normalized by $E_{\text{pot}}(0)=h\times 120$~Hz. Because of the trap anisotropy, the quadrupole mode is excited after a few oscillation, and the potential energy displays an amplitude modulation. Solid line is the sinusoidal fit with two frequencies, $\omega_B=2\pi\times 20.1(2)$~Hz and $\omega_Q=2\pi\times 14.2(6)$~Hz. The measured frequencies are closed to the collective excitation modes, breathing mode ($\omega_B=2\omega_r$) and quadrupole mode ($\omega_Q=\sqrt{2}\omega_r$), in an isotropic trap with the mean trap frequency ($\omega_r=2\pi\times9$~Hz). Each data point averages over 3 measurements, and the error bars mean the one standard deviation.
\label{Breathingmode}}
\end{figure}

 We excited the breathing mode by applying an RF-pulse to the polar condensate under  5.6~G of magnetic field, where the quadratic Zeeman shift ($q=19$~kHz) was sufficiently large so that all atoms were transferred to the $|m_F=1\rangle$ state. Assuming that the condensate profile was preserved during the evolution, we calculated the potential energy using the Thomas-Fermi fit to the condensates, and plot the time evolution of the normalized potential energy $E_{\text{pot}}(t)/E_{\text{pot}}(0)$ in Fig.~6(c). The potential energy oscillates periodically over the hold time, and we observe an additional weak amplitude modulation with a frequency of $\omega_Q=14$~Hz. This can be attributed to the trap anisotropy ($\omega_x/\omega_y=0.8$), which couples a radial monopole mode to a quadrupole mode. As a result, the above simple relations need corrections, and numerical studies on the superfluid hydrodynamics will be required to precisely measure the scattering length ratio after the interaction quench.
 
  In this study, we empirically estimated the scattering length ratio by fitting the first three oscillations to a single cosine function in Eq.~(2), where the quadrupole mode was not significantly developed. The fitted oscillation frequency is $\omega_B=2\pi\times~21.6(6)$~Hz, which is comparable to the twice mean trap frequency, $2\omega_r=\omega_x+\omega_y=2\pi\times18$~Hz. To measure the scattering length ratio, we take the relation of the normalized offset Eq.~(4) because the oscillation amplitude gradually decreases in the first three cycles, and obtain the ratio $\gamma=0.27(6)$. The result is consistent with the estimation that is obtained by comparing spin polarized condensates size and close to the theoretical estimation $a_{f=2}/a_{f=0}=0.28$, which might indicate the correction term for the trap anisotropy could be small in our experiment.

\section{Summary and Outlook}
We have prepared the Bose-Einstein condensates of $^7$Li atoms in a weak magnetic field and studied the non-equilibrium spin dynamics of ferromagnetic spinor BECs. To measure the scattering length difference among $F=1$ hyperfine states, we investigated the quantum phase transition from the P phase to the EPF phase and coherent spinor dynamics under various magnetic fields. Both experiments gave similar results for scattering length difference, $a_{f=2}-a_{f=0}=-18(3)~a_{\text{B}}$. Additionally, we obtained the scattering length ratio $a_{f=2}/a_{f=0}=0.27(6)$ by studying the 2D breathing mode dynamics, which was induced by a spin-flip transition. Taking all measurements together, we determined the scattering length for each spin channel to be $a_{f=2}=7(2)~a_{\text{B}}$ and $a_{f=0}=25(5)~a_{\text{B}}$, respectively.

The results demonstrate a strongly ferromagnetic spinor condensate of $^7$Li atoms, where the spin interaction is as large as 46$\%$ of the density-density interaction, and can be extended in many ways. It allows us to study the Ising ferromagnetic instability and other symmetry breaking phase transitions above the Bose-Einstein condensation temperature~\cite{Natu2011}, and investigate the relationship between spin and mass superfluidity near the quantum critical point $q_{c'}=0$~\cite{Armaitis2017,Sonin2018}. Moreover, given the fast time scale for coarsening dynamics and with the long life time of the condensates, we could explore long time thermalization processes and investigate universal behavior in non-equilibrium quantum systems after compensating the field gradient~\cite{Williamson2016,Williamson2017,Fujimoto2018,Maximilian2018,Schmied2019}. 


\begin{acknowledgments}
The authors thank  Junhyeok Hur and Haejun Jung for discussion and critical reading of the manuscript. This work was supported by National Research Foundation of Korea (NRF) Grant under project number 2019M3E4A1080401 and  2020R1C1C1010863.
\end{acknowledgments}

\newcommand{\beginsupplement}{%
        \setcounter{table}{0}
        \renewcommand{\thetable}{S\arabic{table}}%
        \setcounter{figure}{0}
        \renewcommand{\thefigure}{S\arabic{figure}}%
     }
     
\section{Appendix}
\beginsupplement

\subsection{Evaporation cooling to BEC}

\begin{figure}
\includegraphics[width=0.85\columnwidth]{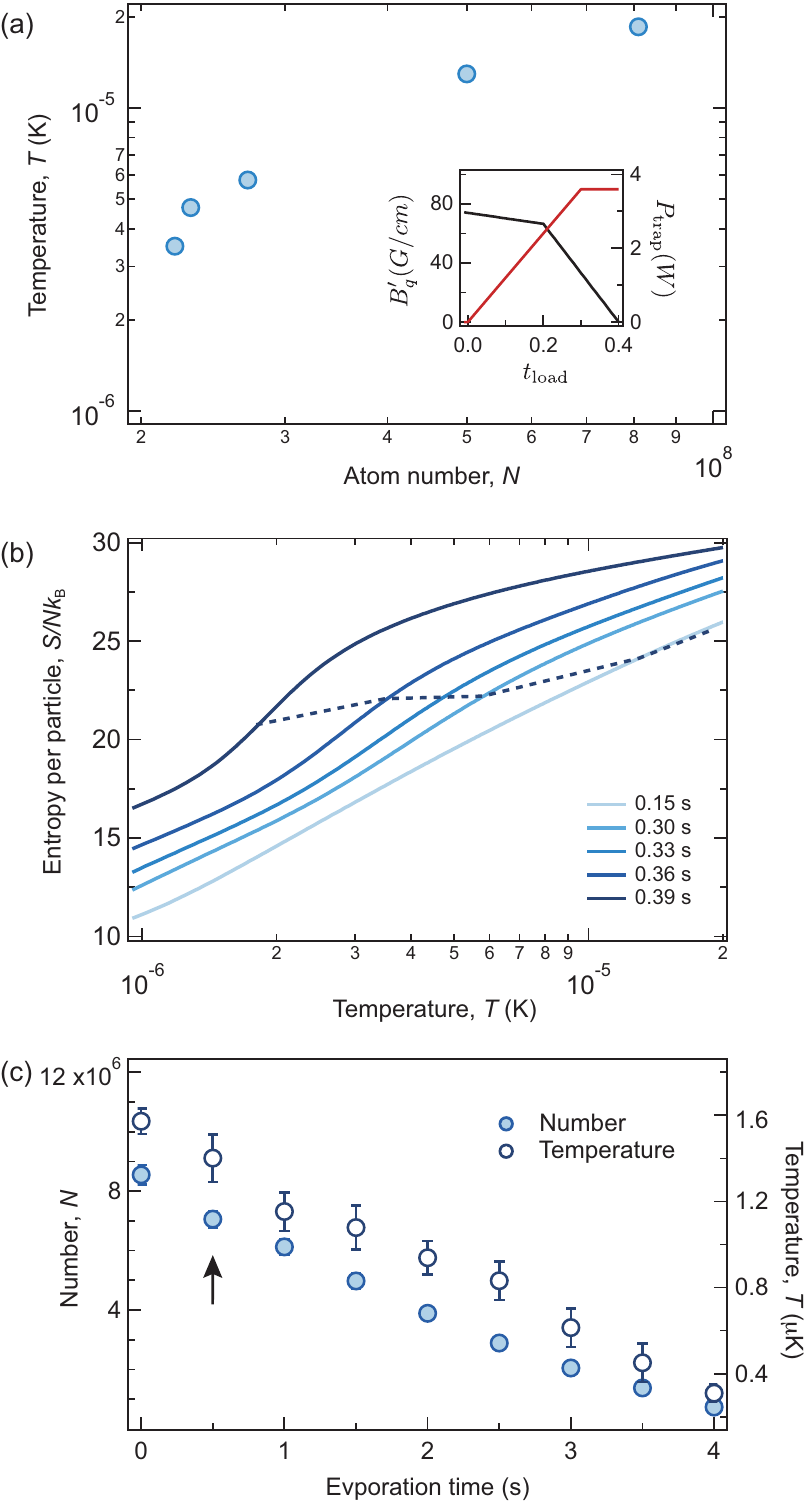}
\caption{Loading and cooling thermal gases to generate Bose-Einstein condensates. (a) Trajectory of temperature ($T$) and number  ($N$) of cold atoms during the optical trap transfer. The inset shows experimental sequence of switching potential from the quadrupole trap to the qausi-2D optical trap. After the transfer, we are able to have 10$\%$ of atoms in the magnetic trap. (b) Estimated entropy per particle during the transfer process. Dashed line marks the  measured entropy trajectory. (c). Atom number and temperature as a function of evaporation time. The arrow indicates the birth of Bose-Einstein condensation. All data points are measured over 3 independent realizations, and the error bars are the 1 s.d. fluctuations of the data.
\label{Transfer}}
\end{figure}

In this section, we describe in more detail the experimental processes used to generate the BECs without using Feshbach resonance. To obtain large atom number condensates, we first performed microwave evaporation cooling in a plugged quadrupole magnetic trap with upper hyperfine spin state $|F=2,m_F=2\rangle$, and then transferred the cold atoms to a quasi-2D optical dipole trap. The quasi-2D optical trap is made of a single optical sheet with a 1070~nm wavelength. Its $1/e^2$ beam waist is $11~\mu$m ($0.95$~mm) in the $z (y)-$direction~\cite{Clade2009}. The optical trap is displaced by 200~$\mu$m below the zero-field position of the quadrupole trap so that it does not interfere with the optical plugging. After cooling the atoms to 20~$\mu$K, we gradually turned on the optical trap, producing potential depth $U_0=10~\mu{}$K in 300~ms, and ramped down the field gradient $B_q'$ to zero in 400~ms (Fig. S1, inset). The microwave was swept from 807~MHz to 804~MHz during this process. 

Right after the transfer, thermal gases with $N=8.5\times10^6$ number of atoms were prepared and reached thermal equilibrium at $2.3~\mu{}$K. We note that the temperature in the dipole trap was already very close to the condensate critical temperature $T_c=1.7~\mu{}$K with the peak phase space density was order of unity. Indeed, the condensates were observed after 0.5~s of evaporation cooling by lowering the optical power. To have better understandings of the transfer process, we calculated the entropy per particle ($S/Nk_B$) by measuring the atom number and temperature at various loading time [Fig.~S1 (a)]. The analysis was done with the effective potential formed by the quadrupole magnetic trap and the quasi-2D optical trap~\cite{Lin2009}. The $S/Nk_B$ curves shows we are able to cool down the atoms to $3.5~\mu{}$K after an adiabatic transfer, which is attributed to large trap volume of the quasi-2D potential.

The final evaporation was done by lowering the trap depth in 5~s with an exponential time constant of 2~s. To maximize intraspecies collision rate among the $F=1$ hyperfine state, we used the lower hyperfine spin state $|F=1,m_F=0\rangle$, which was prepared by applying double Landau-Zener sweeps. At the beginning of the evaporation we should keep the magnetic field to 20~G in order to suppress thermal population in the other spin states. Fig.~S1(c) displays the total atom number ($N$) and temperature ($T$) during the evaporation process. 

\subsection{Longitudinal magnetization}

\begin{figure}
\includegraphics[width=0.9\columnwidth]{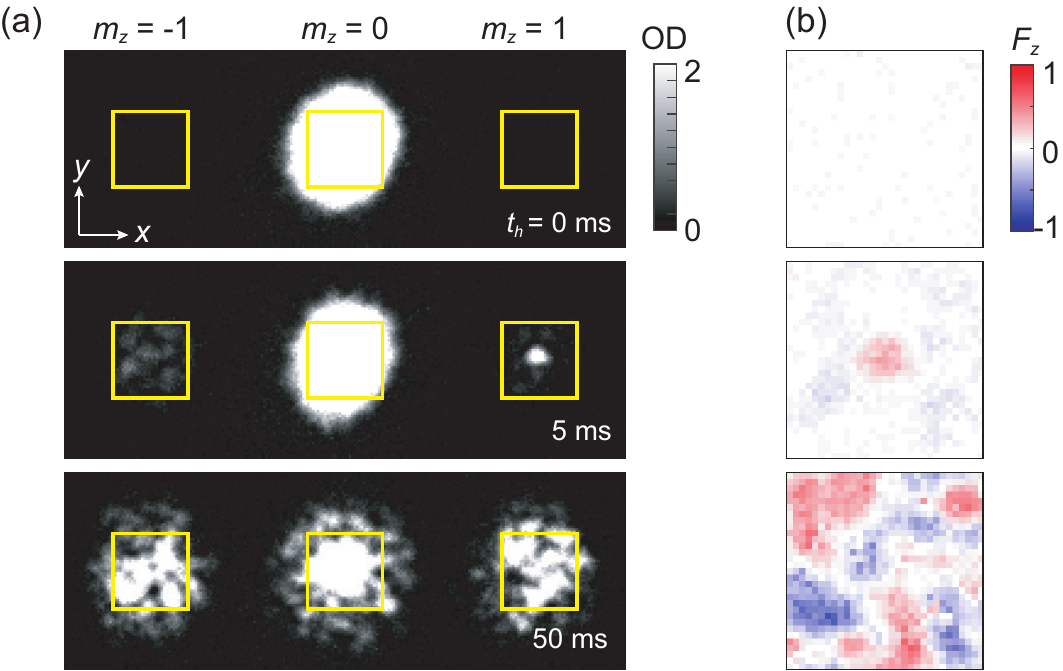}
\caption{Spin dynamics after quenching the magnetic field. (a) Stern-Gerlach spin-separated absorption images at various hold time, $t_h$. The yellow square boxes ($120~\mu{}$m$\times120~\mu{}$m) are the central region of interest for each spin state. (b) Reconstructed images of $\langle F_z(r)\rangle$ for each hold time. 
\label{SGimages}}
\end{figure}
We analyze magnetization along $z$-axis by taking spin-separated absorption images [Fig.~S2]. We take central region of condensate for each spin state to minimize the density inhomogeneous effect. The longitudinal magnetization is calculated by 
\begin{equation}
\langle F_z\rangle=\frac{n_1-n_{-1}}{n_1+n_0+n_{-1}},
\end{equation} where the $n_{j}$ is the atomic density of $|m_z=j\rangle$ spin state. 



\newpage
\bibliography{SMD}
\end{document}